\documentclass{aa}
\usepackage{graphics}

\begin{document}

   \thesaurus{1    % A&A Letters
              (03.19.3 % Space vehicles : instruments
			   08.05.2 % Stars : emission-line, Be
			   08.16.4 % Stars : AGB and post-AGB
			   11.13.1 % (Galaxies:) Magellanic Clouds
			   13.09.6)} % Infrared : stars

   \title{New infrared object in the field of the SMC cluster NGC 330\thanks 
   {Based on observations with ISO, an ESA project with instruments funded by 
   ESA Member States (especially the PI countries: France, Germany, the 
   Netherlands and the United Kingdom) and with the participation of ISAS 
   and NASA.}  
   }

   \author{A. Ku\v{c}inskas
         \inst{1,2,3} \thanks{Research Fellow of the Japan Society for the Promotion of Science}
         \and
         V. Vansevi\v{c}ius
         \inst{4}
	     \and
	     M. Sauvage
         \inst{5}
		 \and
         T. Tanab\'{e}
         \inst{6}
		 }

   \offprints{A. Ku\v{c}inskas}

   \institute{National Astronomical Observatory, Tokyo, 181-8588, Japan
         \and
         Institute of Theoretical Physics and Astronomy, Go\v {s}tauto 12, Vilnius 2600, Lithuania
	     \and
	     Institute of Material Research and Applied Science, Vilnius University, \v {C}iurlionio 9, Vilnius 2009, Lithuania
	     \and
	     Institute of Physics, Go\v {s}tauto 12, Vilnius 2600, Lithuania
	     \and
	     CEA/DSM/DAPNIA/Service d'Astrophys. C. E. Saclay, F-91191 Gif-sur-Yvette Cedex, France
	     \and
	     Institute of Astronomy, School of Science, The University of Tokyo, Tokyo, 181-8588, Japan
	     }

  \date{Received : date; accepted : date}

   \maketitle

   \begin{abstract}
   
   We report ISO (Infrared Space Observatory) observations of a new
   infrared source discovered in the vicinity of the young populous 
   cluster NGC 330 in the Small Magellanic Cloud. The object was observed 
   with ISOCAM at 4.5, 6.75 and 11.5 $\mu$m and shows a prominent mid-infrared 
   excess, indicating the presence of a dust shell. The available observations of 
   the optical counterpart together with the mid-infrared ISOCAM data suggest 
   that this object is most likely a post-AGB star, or a Be supergiant. Cluster 
   membership and candidate evolutionary scenarios are discussed briefly.


   \end{abstract}

%
%________________________________________________________________

\section{Introduction}

NGC 330 is a young ($\sim$10-50 Myr, Chiosi et al. 1995, Cassatella et al. 
1996; Keller et al. 1999b) populous cluster in the Small Magellanic Cloud 
(SMC). An intriguing property of this cluster is its richness in Be star 
content (Grebel et al. 1992; Keller et al. 1999c). The Be phenomenon is 
observed in objects with very different evolutionary states, such as 
classical Be stars, Herbig Ae/Be objects, Be supergiants, symbiotic stars 
or post-AGB objects (see e.g. Zickgraf, 1998). Since the age of NGC 330 
is $\le$ 50 Myrs, some of the Be stars `observed in this cluster can indeed 
be classical Be stars, Be supergiants or even Herbig Ae/Be stars. On the 
other hand, stellar evolution theory predicts that at the age of $\sim$50 
Myr high mass stars ($M_{\star}>7 M_{\odot}$) should already reach the AGB 
stage (Fagotto et al. 1994). Since the lifetime of such massive objects on 
the AGB is very short, possibly less than a few $10^6$ years (Bl\" {o}cker, 
1995), some of the objects showing the Be phenomenon in NGC 330 can be thus 
expected to be massive post-AGB stars. 

Additional information helping to distinguish between these different 
evolutionary groups of Be stars can be obtained from the observations 
in the infrared. Except possibly for the classical Be stars, strong emission 
at infrared wavelengths is typical for most objects showing the Be phenomenon, 
including Herbig Ae/Be stars, Be supergiants, post-AGB objects and symbiotic 
stars. Each of these groups can be characterized by different circumstellar 
properties, such as the dust column density, dust temperature and composition, 
and so forth (Waters et al. 1998). Indeed, most of these quantities can be 
constrained from mid-IR observations.

In this work we present the ISOCAM observations of a new infrared source, 
which is the most prominent in the field of NGC 330 at mid-IR wavelengths. 
This object appears to be a strong H$\alpha$ emission source and thus can 
represent a possible Be star candidate. Employing ISOCAM observations and 
data available from the literature we discuss the properties of this object 
and consider several alternatives for constraining its evolutionary status.

%__________________________________________________________________

\section{Observations and results}

The new mid-infrared object (MIR1) was discovered during the raster imaging 
observations of the populous cluster NGC 330 with the ISOCAM (Cesarsky 
et al. 1996) on board the ISO satellite (Kessler et al. 1996). Observations 
were made on May 22, 1997 using the broad-band CAM filters LW1, LW2 and LW10, 
corresponding to the effective wavelengths of 4.5, 6.75 and 11.5 $\mu$m, 
respectively. The raster mode was 5 $\times$ 5, with a raster step size equal to 
8 pixels (24$^{\prime\prime}$) and a pixel field of view (PFOV) of 
3$^{\prime\prime}$. The fundamental integration time was set to $t_{\rm int}$ 
= 2.1 sec, with a total number of about 15 exposures per single raster position. 
ISOCAM data were reduced using the CAM Interactive Analysis software (CIA 
version 3)\footnote{The ISOCAM data presented in this paper was analyzed 
using ``CIA", a joint development by the ESA Astrophysics Division and the 
ISOCAM Consortium led by the ISOCAM PI, C. Cesarsky, Direction des Sciences 
de la Mati\`{e}re, C.E.A., France.} and the photometry was performed with 
the IRAF APPHOT package. 

The measured mid-IR fluxes, together with the optical photometry collected 
from the literature, are given in Table 1. The ISOCAM flux errors given in 
Table 1 are formal APPHOT errors. The absolute photometric uncertainty of 
the ISOCAM measurements is estimated to be less than 20\% (Biviano, 1998).

The optical counterpart of the infrared source was identified from the 
instrumental coordinates of MIR1 which were derived with respect to the 
positions of 8 field stars on LW2 CAM frame (identification accuracy 
is $\sim$ 1$^{\prime\prime}$). The identification chart of MIR1 is given 
in Fig. 1.

\begin{table}[t]
\caption[]{Optical and mid-infrared photometry of MIR1}

\begin{tabular}{cccccc}
\hline
 $B$  &  $V$  &  $I$   & LW1    & LW2    & LW10   \\
 {\rm mag} & {\rm mag} & {\rm mag} & ({\rm mJy}) & ({\rm mJy}) & ({\rm mJy}) \\
\hline
 $17.63^{1}$ & $17.13^{1}$ & $16.32^{1}$ & $4.7 \pm 1.4$ & $12.6 \pm 0.4$ & $32.0 \pm 0.5$ \\
\hline
\end{tabular}

\begin{list}{}{}
\item[$^{1}$] optical data are averages from the following sources : 
	Balona 1992;
	Keller et al. 1999c;
	Sebo \& Wood 1994;
	Vallenari et al. 1994.
\end{list}
\end{table}

%%% Figure 1 %%%

\begin{figure}
\resizebox{\hsize}{!}{\includegraphics{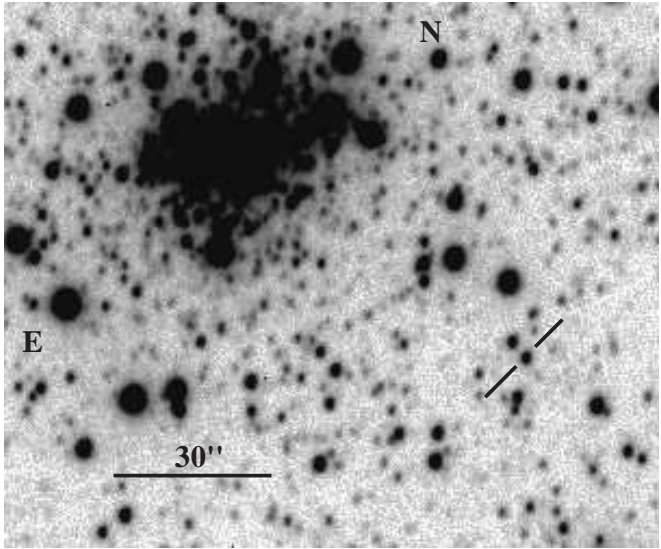}}
\caption{Identification chart of MIR1 in I-band (north is up and east is left). 
	The frame also covers the central part of SMC cluster NGC 330.}
\end{figure}

%__________________________________________________________________

\section{Discussion}

The optical counterpart of MIR1 appears to be the variable star 224 discovered 
by Balona (1992). During the observing run of six nights when it was monitored 
by Balona, it faded by 0.2 mag and was distinctly variable within a night. 
Although periods around the 1 day expected for a Cepheid were indicated, no 
period gave a satisfactory fit to the data so the observed scatter and red 
color led Balona to suspect that it may be a double mode Cepheid on the red 
edge of the instability strip. Independent observations of NGC 330 (Sebo \& 
Wood 1994) made over a 4 year period verified the variability of MIR1 (their 
star 515V)with a $\Delta V\approx0.5$ and $\Delta I\approx0.4$, but again no 
regular period was evident. Strikingly, the average V magnitude over six days 
(17.12; Balona, 1992) is very similar to the average V magnitude over $\sim$4 
years (17.17; Sebo \& Wood, 1994).

The optical counterpart of MIR1 was found to be a strong H$\alpha$ source. 
Observations in the narrow-band ($\Delta\lambda = 1.5$nm) H$\alpha$ 
filter showed that this object (star 485, Keller et al. 1999c) was the 
second strongest H$\alpha$ emitter in the field of NGC 330 after the 
planetary nebula L305. This object is also listed in the SMC H$\alpha$ 
source catalog of Meyssonnier \& Azzopardi (1993) as object 906.

The strong H$\alpha$ emission and the prominent mid-IR excess are difficult 
to assess within the evolutionary scenario of a classical Cepheid. Indeed it 
is possible that this object is a binary system, however the discussion of 
this possibility in the view of the scarce observational facts seems rather 
premature. The $({\rm H}\alpha - R)$ color index is much larger in MIR1 than 
in any classical Be star in NGC 330 (Keller et al. 1999c), which, together with 
a strong mid-IR excess, indicates that MIR1 is unlikely to be a classical Be 
star. Therefore, we will further concentrate on the Be supergiant, Herbig 
Ae/Be and post-AGB star scenarios instead.

\subsection{Be supergiant and Herbig Ae/Be star scenarios}

One of the possible alternatives for constraining the evolutionary status 
of MIR1 is a Be supergiant scenario. This is indeed supported by the existence 
of H$\alpha$ emission, which is typical to all types of Be stars. Spectral 
observations of MIR1 obtained by Keller (1999a) confirm that this object is 
a very strong H$\alpha$ emitter; the spectrum clearly shows H$\alpha$ line 
though no H$\gamma$ or higher.

Although the observed optical color indices of MIR1 are distinctively different 
from those of Be supergiants in the Magellanic Clouds (Zickgraf et al. 1992), 
this may be a consequence of the interstellar or circumstellar reddening. A 
dereddening procedure employing the reddening-free $Q$ parameter yields $Q_{BVI}
\approx-0.09$ (calculated assuming the standard excess ratio) which indicates 
that the spectral type of this object (depending on the luminosity class) should 
be O8-B2. Taking B0 as a representative of these values, one obtains $V_{0}\sim
14.60$, $(B-V)_{0}\sim-0.30$ and $(V-I)_{0}\sim-0.27$, $A_V\sim2.5$ and, using 
the SMC distance modulus of 18.9, $M_V\sim-4.3$. Assuming that the bolometric 
correction for the spectral type B0 is $BC_V\sim-2.5$, we derive $M_{\rm bol}\sim-6.8$. 
Taking into account the errors of the spectral type determination (which set a 
range of possible $T_{\rm eff}$ between $20\,000-34\,000$ K), the obtained 
$T_{\rm eff}$ and $M_{\rm bol}$ are indeed comparable with those of Be supergiants 
in the MCs (cf. Zickgraf et al. 1992). Keller et al. (1999b) show a HR diagram of 
the cluster from the HST data and the Be stars at the cluster turnoff have 
$T_{\rm eff}\sim16\,000$ and $M_{\rm bol}\sim-5.8$. They also have one Be 
star (B13) like a blue straggler with $T_{\rm eff}\approx32\,000 K$ and 
$M_{\rm bol}\sim-6.5$. These temperatures and luminosities are similar to the ones 
obtained for MIR1. The derived $A_V\sim2.5$, however is much higher than the average 
in the field of NGC 330 (which measures the range from $E(B-V)=0.03$ derived by 
Carney et al. (1985) to $E(B-V)=0.12$ obtained by Bessell, 1991), and therefore 
indicates a significant circumstellar extinction.

Indeed, the spectral energy distribution of MIR1 shows a strong mid-IR excess 
(Fig. 2). The estimate of the ratio of ISO LW10 band flux over the V band flux 
in MIR1 yields $F_{12}/F_V\sim3.3$. This is comparable with the $F_{12}/F_V\sim5.4$ 
observed in a 'representative' Be supergiant GG Car (Waters et al. 1998) and thus 
could be viewed as an additional argument supporting the Be supergiant scenario.

Employing theoretical evolutionary tracks of Fagotto et al. (1994) and making 
use of the derived $T_{\rm eff}$ and $M_{\rm bol}$ we obtain a stellar mass of 
$M_{\star}\sim15-20 M_{\odot}$ and the age of 8-14 Myr. The derived age of MIR1 
is comparable with the cluster's age (10-20 Myr, Cassatella et al. 1996), 
suggesting that the candidate Be supergiant could be a cluster member.

Prominent mid-IR excesses are also common in Galactic Herbig Ae/Be stars with 
cool circumstellar shells (group II objects, see Hillenbrand et al. 1992). However, 
Herbig Ae/Be scenario seems rather unlikely for the case of MIR1. First, the 
available observations of NGC 330 do not show any evidence for the ongoing star 
formation in the field of NGC 330. Second, although some Galactic Herbig Ae/Be 
stars are observed as isolated objects, they are usually low-mass stars (cf. 
Hillenbrand et al. 1995) and therefore the high mass of the possible Herbig Ae/Be 
candidate ($\ge25$ $M_{\odot}$) inferred from the dereddened photometry of 
MIR1 and the SMC distance modulus rules out this possibility too.

%%% Figure 2 %%%

\begin{figure}
\resizebox{\hsize}{!}{\includegraphics{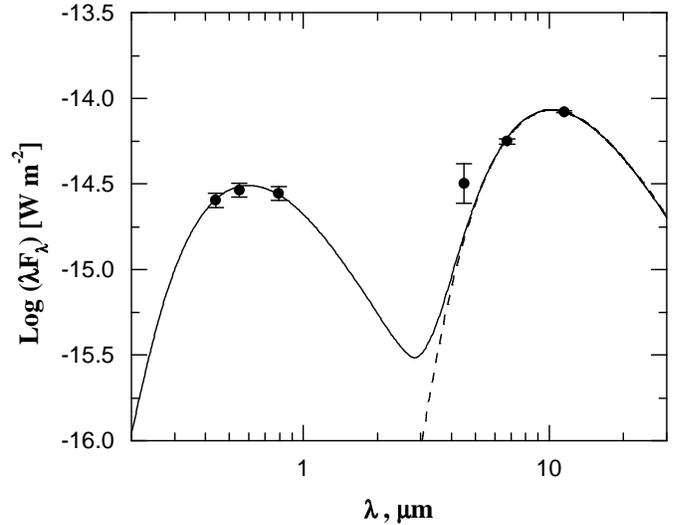}}
\caption{Spectral energy distribution of MIR1, constructed from optical 
	photometry and ISO data (Table 1). Error bars of the mid-IR data 
	are formal IRAF/APPHOT errors. Solid line shows two-blackbody 
	fit to the optical and mid-IR data ($T_{\rm BB_1}=6200$ K and 
	$T_{\rm BB_2}=360$ 
	K); dashed line indicates $T_{BB}=360$ K fit to the mid-IR ISOCAM data 
	used to estimate the infrared luminosity $L_{\rm IR}$ (see text for details).}
\end{figure}

\subsection{Post-AGB star scenario}

Post-AGB stars have been long recognized as one of the evolutionary groups 
showing the Be phenomenon. Indeed, strong H$\alpha$ emission is typical for 
most post-AGB objects and thus the existence of H$\alpha$ emission in MIR1 
works in favor of this scenario too. 

Most of the post-AGB objects show a double-peaked spectral energy distributions 
(e.g., Kwok, 1993; Zhang \& Kwok, 1991), similar to the one observed in MIR1 
(Fig. 2). A simple estimate of the infrared luminosity obtained from the 
blackbody fit to the ISOCAM data yields $L_{\rm IR}\sim1300 L_{\odot}$ with 
a blackbody dust temperature $T_{\rm d}=360$ K. The estimate of the dust mass 
in the circumstellar shell, $M_{\rm d}$, can be made then using the following 
expression (Gurzadyan, 1997) :

\begin{equation}
{{M_{\rm d}}\over {M_{\odot}}} = 9.21\,{T^{-4}_{\rm d}}\,{{L_{\rm IR}}\over {L_{\odot}}}
\end{equation}

\noindent where $T_{\rm d}$ and $L_{\rm IR}$ are the dust temperature and the 
infrared luminosity, respectively. Taking the $L_{\rm IR}$ and $T_{\rm d}$ values 
derived above, one obtains $M_{\rm d}\sim7\times10^{-7} M_{\odot}$, which is 
comparable with the dust masses typical for the post-AGB objects (e.g., Pottasch 
\& Parthasarathy, 1988). Two facts should be noted, however. Firstly, the 
obtained blackbody dust temperature ($T_{\rm d}=360$ K) can be considerably 
overestimated, since its derivation relies on the mid-IR data only and 
does not take into account any information about the dust radiation at 
longer wavelengths. Secondly, at the dust temperatures typical for the 
post-AGB objects, a large fraction of infrared flux is emitted at 
wavelengths longer than $\sim12 \mu$m and thus $L_{\rm IR}$ can be 
considerably higher than the presently derived value. Therefore, the 
obtained estimate of $M_{\rm d}$ indicates only a lower limit for the dust 
mass in MIR1.

The upper limit for the effective temperature of the central star of the 
possible post-AGB object can be inferred from the following considerations. 
If MIR1 is assumed to be a normal planetary nebula (i.e., past the PPN stage), 
the effective temperature of the central star should be at least $T_{\rm eff}
\sim30\,000$ K and the observed $(B-V)=0.50$ would indicate a considerable 
circumstellar extinction. Indeed, the central star with $T_{\rm eff}\sim30\,000$ 
K should have $(B-V)_0\sim-0.30$, and hence the $E(B-V)\sim0.8$, that is, 
$A_V\sim2.5$ and $M_V\sim-4.0$. Assuming that the bolometric correction is 
$BC_V\sim-3.0$ one obtains $M_{\rm bol}\sim-7.0$, which is very close to the 
classical luminosity limit for the post-AGB stars ($M_{\rm bol}\sim-7.2$, e.g. 
Shaw \& Kaler, 1989). Thus we conclude, that the classical luminosity limit 
for the post-AGB objects sets the upper limit for the effective temperature 
of the central star at about 30\,000 K.

The lower limit for the effective temperature of MIR1 can be constrained from 
the observed SED. The two-blackbody fit to the optical and mid-IR data (see 
Fig. 2) gives a lower limit estimate of the total luminosity of MIR1, $L_{\rm tot}
\sim1800 L_{\odot}$. Using a simple iteration procedure one can obtain the 
$(B-V)_0$, and therefore $T_{\rm eff}$, which would produce the observed 
$L_{\rm tot}$ with the observed $M_V\sim-1.8$. Such procedure yields $(B-V)_0
\sim-0.14$, $A_V\sim2.0$, and $T_{\rm eff}\sim14\,000$ K, seting this as a 
lower limit for the effective temperature of the central star.

The obtained temperature range suggests that MIR1 can be a good proto-planetary 
nebula (PPN) candidate. This is reinforced by the fact, that the infrared to 
the total luminosity ratio in MIR1 is $L_{\rm IR}/L_{\rm tot}\sim0.7$, which is 
considerably higher than the value typical for the planetary nebulae ($L_{\rm IR}/
L_{\rm tot}\sim0.3$, see e.g. Pottasch, 1997). Since the presently estimated 
total luminosity of MIR1 is only $L_{\rm tot}\sim1800 L_{\odot}$, it is rather 
unlikely that this object could be a high mass post-AGB star belonging to NGC 
330; instead, it is probably a low mass field star. However, the mass and thus 
the evolutionary status of the possible PPN can not be constrained precisely 
yet. Therefore the tighter constraints on this scenario should be set by future 
optical spectroscopy of MIR1, which would provide additional information both 
about the central star and the nebula.

%__________________________________________________________________

\section{Conclusions}

We present the ISOCAM observations of a new infrared object in the field 
of the young populous cluster NGC 330 in the Small Magellanic Cloud. The 
discovered IR object which has been previously identified as a variable star 
shows a prominent mid-IR excess, indicating the presence of a dust shell. 
This along with the strong H$\alpha$ emission makes the suggestion that this 
object may be a low-mass Cepheid rather unlikely. We suggest instead that 
this object may be a low mass field post-AGB star in a proto-planetary nebula 
stage or a Be supergiant belonging to cluster NGC 330. In both cases the expected 
optical extinction is relatively high ($A_V\sim2.0-2.5$). We also can not 
reject the possibility that this object may be an isolated Herbig Ae/Be star. 
Unfortunately, presently available observations do not allow to distinguish 
between these scenarios clearly. Neither the basic stellar parameters of the 
detected object, nor the physical conditions in the H$\alpha$ and dust emitting 
regions can be constrained precisely yet. Therefore, further photometric 
observations in the UV, near-IR and far-IR, and especially optical spectroscopy 
would be a highly desirable second step towards clarifying the nature of MIR1.

\acknowledgements 

We thank Saulius Raudeli\={u}nas and Mudumba Parthasarathy for productive and
stimulating discussions, Stefan Keller for providing information about 
spectral observations of MIR1 prior to publication, and the referee, Michael 
Bessell, for the valuable comments and suggestions. This research was supported 
in part by grant-in-aids for Scientific Research (C) and for International 
Scientific Research (Joint Research) from the Ministry of Education, Science, 
Sports and Culture in Japan.

\end{document}